\documentclass[a4paper,11pt]{article}
\usepackage{pos}
\usepackage{graphicx}
\usepackage[caption=false]{subfig}
\usepackage{float}
\usepackage{wrapfig}
\usepackage{multirow}
\usepackage{longtable}
\usepackage{url}
\usepackage{bm}
\usepackage{mathrsfs}
\usepackage{amsmath}
\usepackage{amsfonts}
\usepackage{color}
\usepackage{xspace}
\usepackage{tikz}
\usepackage{slashed}
\usepackage{cancel}
\usepackage{hyperref}
\usepackage{xcolor}
\usepackage{slashed}
\usepackage[symbol]{footmisc}
\renewcommand{\thefootnote}{\fnsymbol{footnote}}
\usepackage[capitalize]{cleveref}
\creflabelformat{equation}{#2\textup{#1}#3}

\title{Deep Learning for the Matrix Element Method}

\author[1]{Matthew Feickert}
\author[2]{Mihir Katare}
\author*[2,3]{Mark Neubauer\footnote[2]{msn@illinois.edu}}
\author[2,3]{Avik Roy}
\affiliation[1]{Department of Physics, University of Wisconsin-Madison, Madison, WI 53706 USA}
\affiliation[2]{Department of Physics, University of Illinois at Urbana-Champaign, Urbana, IL 61801 USA}
\affiliation[3]{National Center for Supercomputing Applications, 1205 Clark St, Urbana, IL 61801, USA}

\abstract{
Extracting scientific results from high-energy collider data involves the comparison of data collected from the experiments with “synthetic” data produced from computationally-intensive simulations. Comparisons of experimental data and predictions from simulations increasingly utilize machine learning (ML) methods to try to overcome these computational challenges and enhance the data analysis. There is increasing awareness about challenges surrounding interpretability of ML models applied to data to explain these models and validate scientific conclusions based upon them. The matrix element (ME) method is a powerful technique for analysis of particle collider data that utilizes an \textit{ab initio} calculation of the approximate probability density function for a collision event to be due to a physics process of interest. The ME method has several unique and desirable features, including (1) not requiring training data since it is an \textit{ab initio} calculation of event probabilities, (2) incorporating all available kinematic information of a hypothesized process, including correlations, without the need for “feature engineering” and (3) a clear physical interpretation in terms of transition probabilities within the framework of quantum field theory. These proceedings briefly describe an application of deep learning that dramatically speeds-up ME method calculations and novel cyberinfrastructure developed to execute ME-based analyses on heterogeneous computing platforms. 
}

\FullConference{%
  41st International Conference on High Energy physics - ICHEP2022\\
  6-13 July, 2022\\
  Bologna, Italy
}

\begin{document}
\maketitle


\section{Introduction}
\label{sec:introduction}
The Matrix Element (ME) Method~\cite{Kondo:1988yd,Fiedler:2010sg,2011arXiv1101.2259V,Elahi:2017ppe} is a powerful technique which can be use for measurements and direct searches for new phenomena. It has been used extensively by collider experiments at the Tevatron for Standard Model (SM) measurements and Higgs boson searches~\cite{Abazov:2004cs,Abulencia:2006ry,Aaltonen:2008mv,Aaltonen:2010cm,Abazov:2009ii,Aaltonen:2009jj} and at the LHC for measurements in the Higgs and top quark sectors of the SM~\cite{Chatrchyan:2012xdj,Chatrchyan:2013mxa,Aad:2014eva,Khachatryan:2015tzo, Khachatryan:2015ila,Aad:2015gra,Aad:2015upn}.
The ME method is based on \textit{ab initio} calculation of the probability density function $\mathcal{P}$ of an event with observed final-state particle momenta ${\bf x}$ to be due to a physics process $\xi$ with theory parameters $\boldsymbol\alpha$.
One can compute $\mathcal{P}_{\xi}({\bf x}|{\boldsymbol\alpha})$ by means of the factorization theorem from the partonic cross-sections of the hard scattering process involving parton momenta ${\bf y}$ and is given by 
\begin{equation}
\mathcal{P}_{\xi}({\bf x}|{\boldsymbol\alpha}) = \frac{1}{\sigma_{\xi}(\boldsymbol\alpha)} \int d\Phi ({\bf y}_{\rm final}) \; dx_1 \; dx_2~\frac{f(x_1)f(x_2)}{2s x_1 x_2} \; |\mathcal{M}_{\xi}({\bf y}|\boldsymbol\alpha)|^2 \; \delta^{4}({\bf y}_{\rm initial}-{\bf y}_{\rm final}) \; W({\bf x}, {\bf y})
\label{eqn:MEProb}
\end{equation}
\noindent where and $x_i$ and ${\bf y}_{{\rm initial}}$ are related by $y_{{\rm initial},i}\equiv \frac{\sqrt{s}}{2}(x_i,0,0,\pm x_i)$, $f(x_i)$ are the parton distribution functions, $\sqrt{s}$ is the collider center-of-mass energy, $\sigma_{\xi}(\boldsymbol\alpha)$ is the total cross section for the process $\xi$ (with $\boldsymbol\alpha$) times the detector acceptance, $d\Phi({\bf y})$ is the phase space density factor, $\mathcal{M}_{\xi}({\bf y}|\boldsymbol\alpha)$ is the matrix element (typically at leading-order (LO)), and $W({\bf x}, {\bf y})$ is the probability density (``transfer function'') that a selected event ${\bf y}$ ends up as a measured event ${\bf x}$. In practice, $W({\bf x}, {\bf y})$ is a simplified version of the full detector response to particle collisions derived from an analysis of fully-simulated events and different choices of $W({\bf x}, {\bf y})$ are studied to probe dependence on $W({\bf x}, {\bf y})$.

One can use \Cref{eqn:MEProb} to analyze data from particle colliders. For measurement of model parameters $\boldsymbol\alpha$, one maximizes the likelihood function for data events $\mathcal{L}(\boldsymbol\alpha)$ given by
\begin{equation}
\mathcal{L}(\boldsymbol\alpha) = \prod_{i} \sum_{k} f_k \mathcal{P}_{\xi_k}({\bf x}_i|{\boldsymbol\alpha})
\label{eqn:LH}
\end{equation}
\noindent where $f_k$ are the fractions of processes $\xi_k$ contributing to the data. For particle searches, one can use Bayes' Theorem~\cite{Bayes01011763} to compute, for a hypothesized signal $S$, the probability $p(S|{\bf x})$ given by
\begin{equation}
p(S|{\bf x}) = \frac{\displaystyle\sum_{i} \beta_{S_i} \mathcal{P}_{S_i}({\bf x}|\boldsymbol\alpha_{S_i}) }{\displaystyle\sum_{i} \beta_{S_i} \mathcal{P}({\bf x}|\boldsymbol\alpha_{S_i}) + \sum_{j} \beta_{B_j} \mathcal{P}({\bf x}|\boldsymbol\alpha_{B_j})}
\label{eqn:LR}
\end{equation}
\noindent where, $S_i$ and $B_j$, denote considered signal and background processes and $\beta$ are \textit{a priori} expected process fractions.
According to the Neyman-Pearson Lemma~\cite{Neyman289}, \Cref{eqn:LR} is the optimal discriminant for $S$ in the presence of $B$ and can be used to extract a signal fraction in the data.

\section{Challenges and Opportunities}
\label{sec:challenges_opportunities}

The ME method, introduced in Sec.~\ref{sec:introduction} along with several applications, has unique and desirable features, including (1) not requiring training data as an \textit{ab initio} calculation of event probabilities, (2) incorporating all available kinematic information of a hypothesized process, including all final state particle correlations, without the need for ``feature engineering'' and (3) a clear physical interpretation in terms of transition probabilities within the framework of quantum field theory.

The most serious difficulty with the ME method, which has limited its applicability to searches for beyond-the-SM physics and precision measurements at colliders, is that it is \textit{computationally expensive}. Complex final states can take minutes \textit{per event} or more to calculate \Cref{eqn:MEProb} and any practical analysis involves many iterations these calculations. If this limitation could be overcome, as argued in~\cite{HSF-CWP-018,Albertsson:2018maf}, it would enable more widespread use of the ME method and allow for more time for analysis innovation rather than computation.


\section{Containerizing the Workflow}
\label{sec:containerization}

\begin{figure}[h!tb]
\centering
\includegraphics[width=\linewidth]{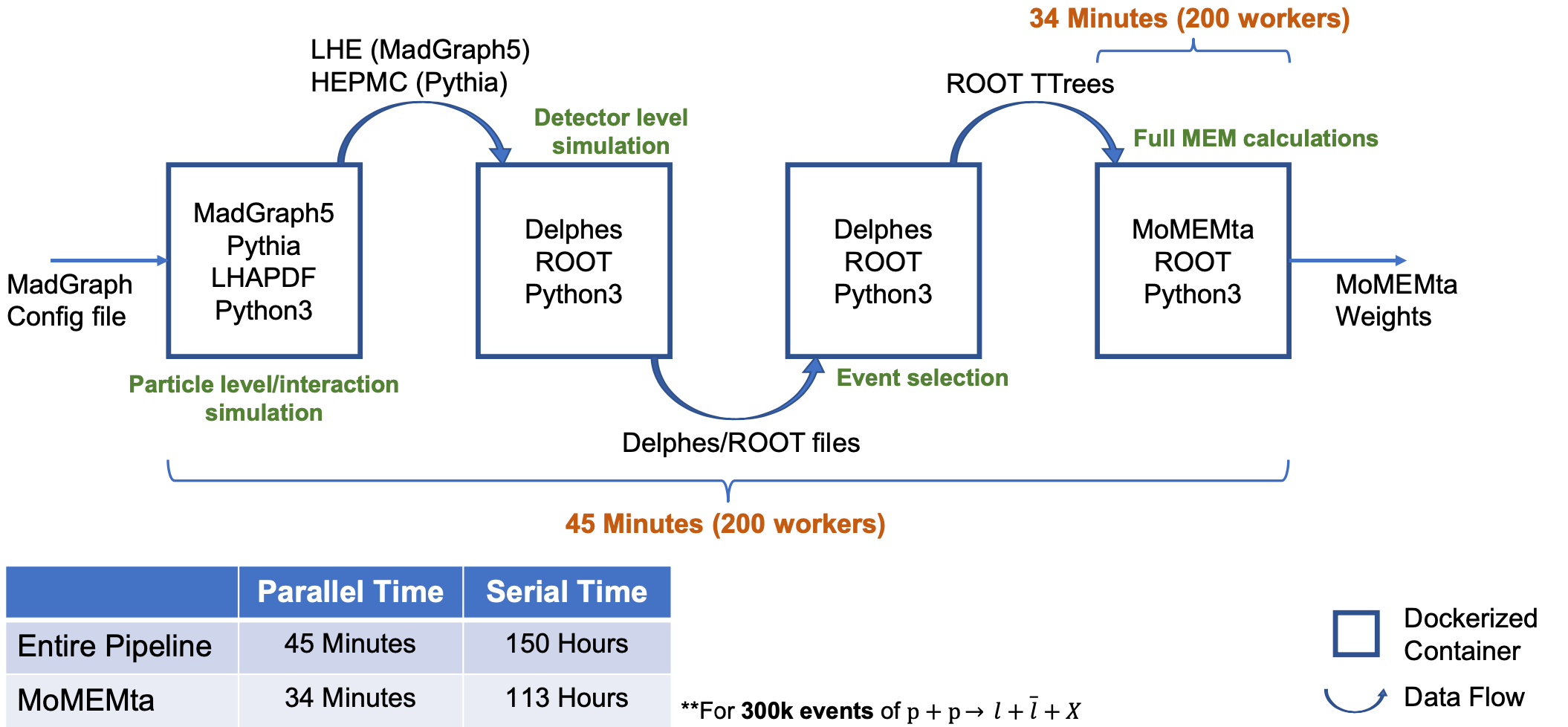}
\caption{Containerized end-to-end simulation+calculation pipeline developed. This pipeline was executed on the Blue Waters Supercomputer at Illinois to simulate and calculate ME probability densities for 300k $pp\to Z/\gamma^* \to \ell^+\ell^- +X$ events in 45 minutes (for comparison, single-threaded execution time is 150 hours)}
\label{fig:MEMcontainers}
\end{figure}

We developed a set of integrated Docker images to provide containers~\cite{MEMcontainer} for an end-to-end MEM-based analysis pipeline shown in Figure~\ref{fig:MEMcontainers}.
This pipeline was executed on the Blue Waters Supercomputer at the University of Illinois to calculate ME probability densities for \cal{O}$(10^7)$ simulated LHC collision events for the $pp\to Z/\gamma^* \to \ell^+\ell^- +X$ (Drell-Yan) process.

\section{Sustainability through Deep Learning}
\label{sec:DeepMEM}

Despite the attractive features of the ME method, the computational burden of the ME method will continue to limit its applicability for practical data analysis without additional innovation. This is especially true when one considers the process of producing a physics publication which involves many selection, sample and systematic iterations for which ME method calculations are required.

\begin{figure}[t!]
\centering
\includegraphics[width=\linewidth]{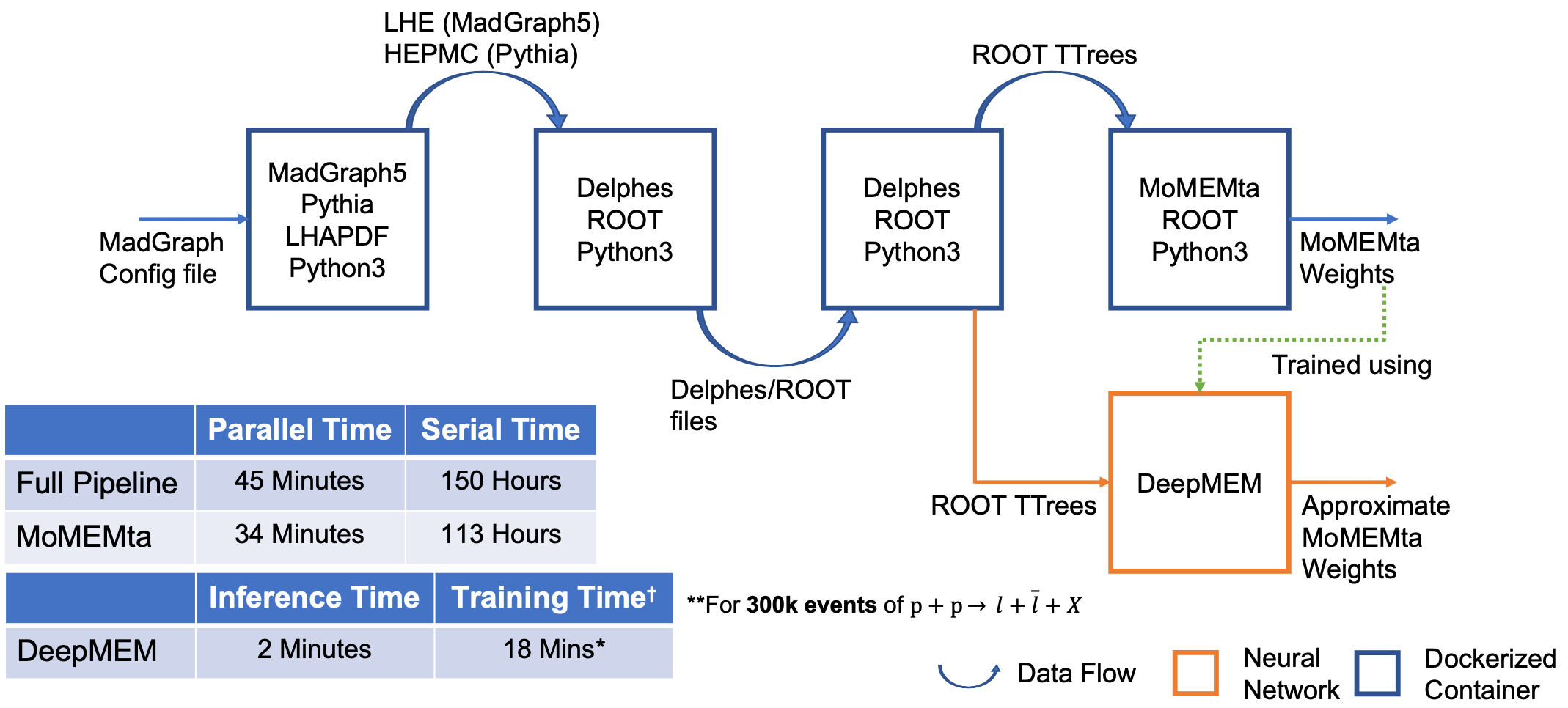}
\caption{Same as Figure~\ref{fig:MEMcontainers} with addition of \texttt{DeepMEM} approximation. After a \textit{one-time} training investment of 18 hours, 300k events were evaluated in 2 minutes with \texttt{DeepMEM} compared to 34 minutes for \texttt{MoMEMta}.}
\label{fig:MEMcontainersDNN}
\end{figure}

As proposed in~\cite{HSF-CWP-018}, ML methods can be used to dramatically speed up the numerical evaluation of \Cref{eqn:MEProb} through evaluation a deep neural network (DNN) trained to approximate \Cref{eqn:MEProb}. The results from~\cite{Bury:2020ewi} treating this as an ME regression problem using a DNN serves as a proof of principle that these techniques are feasible. We refer to our implementation of the DNN approach to MEM regression as \texttt{DeepMEM}~\cite{DeepMEM}. Figure~\ref{fig:MEMcontainersDNN} shows a $\times$17 improvement in the evaluation speed of \Cref{eqn:MEProb} through DNN approximation using \texttt{DeepMEM}.

The model upon which the \texttt{DeepMEM} results just described was based in a fully-connected DNN with five hidden layers each with 200 nodes. This network was trained to learn the result of \Cref{eqn:MEProb} for kinematic inputs of the leptons and jets produced in simulated Drell-Yan events. We split the data 8:1:1 for training, validation and testing purposes, respectively, using a total dataset of $\sim$300k events which were generated using the containerized pipeline described in Sec.~\ref{sec:containerization}.

\begin{wrapfigure}[13]{R}{0.40\textwidth}
\centering
\vspace{-16pt}
\includegraphics[width=\linewidth]{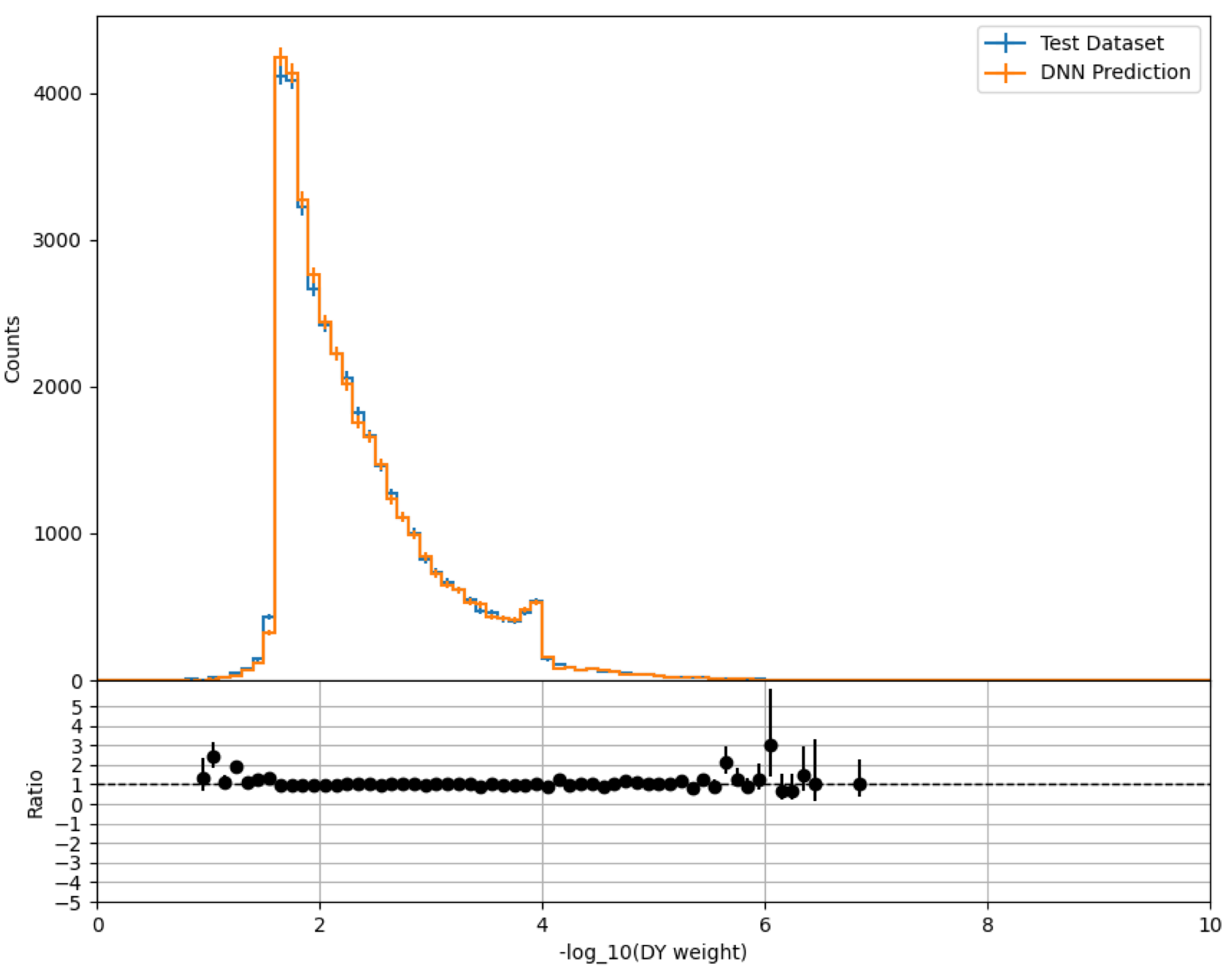}
\vspace{-25pt}
\caption{Comparison of the \texttt{DeepMEM} ME method modeling for simulated Drell-Yan events with calculations from \texttt{MoMEMta}}
\label{fig:DeepMEMmodeling}
\vspace{-32pt}
\end{wrapfigure}
We studied how well the trained model from \texttt{DeepMEM} can approximate the ME probability density calculations from \texttt{MoMEMta}~\cite{Brochet:2018pqf}. The results are shown in Figure~\ref{fig:DeepMEMmodeling} where it is seen that \texttt{DeepMEM} can provide a very accurate approximation of the ME method probability density (\Cref{eqn:MEProb}) as calculated by \texttt{MoMEMta}. We also studied how well the model (without retraining) generalizes to subsets of the kinematic phase space by making cuts on the transverse momentum of the leptons and found similarly good modeling of the \texttt{MoMEMta} calculations, which is a promising result for the method in terms of model generalization.

\vspace*{-0.3cm}
\section{Conclusions and Outlook}
\label{sec:conclusions_outlook}

In this work, we show that a deep neural network model can be trained on ME calculations from \texttt{MoMEMta} to provide an accurate approximation of the MEM-based probability density for Drell-Yan events at a $pp$ collider with detector-level simulation. We show that this model can be used to dramatically speed-up MEM evaluations and demonstrate novel cyberinfrastructure that we developed to execute ME-based analyses on heterogeneous computing platforms including an HPC.

\section*{Acknowledgements}
This work was supported by the U.S. Department of Energy, Office of Science, High Energy Physics, under contract number DE-SC0023365, and by the National Science Foundation under OAC-1841456 and Cooperative Agreement OAC-1836650.


\bibliography{PROC-ICHEP}
\bibliographystyle{JHEP}

\end{document}